\documentclass[prl,superscriptaddress,showpacs,preprint]{revtex4}
\usepackage{graphicx}
\usepackage{dcolumn}
\usepackage{bm}
\usepackage{SIunits}

\newcommand{\mno}{La$_{0.5}$Sr$_{1.5}$MnO$_4$}

\begin{document}

\title{Surface Melting of Electronic Order}

\author{S. B. Wilkins}
\author{X. Liu}
\affiliation{Department of Condensed Matter Physics and Materials Science, Brookhaven National Lab, Upton, New York 11973-5000, USA}
\author{Y. Wakabayashi}
\affiliation{Osaka University, Division of Materials Physics, Graduate School of Engineering Sciences, Osaka 5608531, Japan}
\author{J.-W. Kim}
\author{P. J. Ryan}
\affiliation{Argonne National Laboratory, Advanced Photon Source, 9700 S Cass Ave, Argonne, IL 60439 USA}
\author{J. F. Mitchell}
\affiliation{Argonne National Laboratory, Division of Materials Science, 9700 S Cass Ave, Argonne, IL 60439 USA}
\author{J.P. Hill}
\affiliation{Department of Condensed Matter Physics and Materials Science, Brookhaven National Lab, Upton, New York 11973-5000, USA}

\date{\today}
\begin{abstract}
We report temperature-dependent surface x-ray scattering
studies of the orbital ordered surface in La$_{0.5}$Sr$_{1.5}$MnO$_4$. We find that
the interfacial width of the electronic order grows as the bulk
ordering temperature is approached from below, though the bulk correlation length remains unchanged.
Close to the transition, the surface is so rough that there
is no well-defined electronic surface, despite the presence of 
bulk electronic order, that is the electronic surface has melted. Above
the bulk transition, finite-sized isotropic fluctuations
of orbital order are observed, with a correlation length equal to that of
the electronic surfaces' in-plane correlation length at the transition temperature.
\end{abstract}


\maketitle

Bulk crystals can disorder on warming through one of two scenarios: surface melting or surface freezing. In the first, the surface of the solid melts at a lower temperature than the bulk. This is a common and even useful property of condensed matter and is, for example, what allows a skier to glide across the snow\cite{Dash:2006p2054}.In the second, the bulk melts at a lower temperature than the surface\cite{Ocko:1997p2104,Wu:1993p2062}. An as yet unanswered question is what happens when the order is not that of a crystal lattice, but is rather that of electronic order? We show, in this letter, that orbital order in the manganite \mno\ is an example of electronic surface melting.

The temperature dependence of electronic order at a surface
or interface is of fundamental interest since it can provide
important insights into the nature of electronic interactions,
and in particular the relative strength of the surface and bulk interactions.
Nor is the question entirely academic. Next-generation devices that rely
on electronic order will require understanding and control of electron transport across such
surfacess at finite temperature, and this interfacial
behavior will also have large implications for the properties of any nanoscaled
strongly correlated materials at finite temperature. Thus this is a question of some importance.

Here, we use orbital
order at the surface of a strongly correlated transition metal
oxide as an exemplar system for studying electronic order at a
surface. Before proceding, it is important, to define
what we mean by a surface. The crystallographic surface is
the surface between the solid crystal and the vacuum, as obtained by
cleaving the crystal. The electronic surface (in this case the orbital
surface) is the surface between the orbitally-ordered bulk and an orbitally
disordered region located below the crystallographic surface. 
There has been much work associated with the ground state
behavior of such surfaces, starting with the original suggestion
that ``electronic reconstruction'' can occur at the electronic
order-disorder interface\cite{Okamoto:2004p805}. Surface effects in
electronic order have been seen in such cases as the ferromagnetic surface
of bilayer manganites\cite{Freeland:2005p1290}, the interface between
ferromagnetic and superconducting oxides\cite{Chakhalian:2006p1281},
 the orbital surface of manganites\cite{Wakabayashi:2007p679} and
 the interface between LAO/STO heterostructures\cite{Ohtomo:2004p2072}.
However, to date there has been little work on the evolution
of the electronic surface on warming through the bulk ordering temperature.
The purpose of this work is to redress this imbalance.

We report temperature-dependent surface x-ray scattering
studies of the orbitally ordered surface in \mno\ (LSMO). We find that
the interfacial width of the electronic order grows as the bulk
ordering temperature is approached from below, while the bulk correlation length remains unchanged, suggesting
that the surface starts to roughen while the bulk is well ordered.
Close to, but below, the transition, the surface is so rough that there
is no well-defined electronic surface despite the presence of
bulk electronic order: the electronic surface has melted. Above
the bulk transition,  finite-sized isotropic fluctuations
of orbital order are observed, with a correlation length equal to that of
the surface in-plane correlation length at the transition temperature.
These results have important implications for the behavior of
electronic surface at finite temperatures and we shall discuss these
in view of potential future devices. 

LSMO has been studied extensively
in the bulk, revealing  long range orbital order below
\ensuremath{\sim}240 K, with 3D magnetic order appearing below \ensuremath{\sim}120
K \cite{Sternlieb:1996p2068, Murakami:1998p2059}. This orbital ordering gives rise to concomitant lattice distortions
and thus may be studied utilizing
x-ray scattering techniques. Specifically, by measuring the distribution
of the scattering around the orbital-order superlattice reflection,
one obtains information about the bulk order parameter. Information about
the orbital surface comes from the ``orbital truncation rods''\cite{Wakabayashi:2007p679}. These are rods of
scattering normal to the surface connecting orbital
superlattice peaks and are analogous to crystal truncation
rods \cite{0022-3719-18-35-008, IKRobinsonandDJTweet:1992p2058} which result
from the termination of a 3D crystal by a surface. The intensity
distribution of the orbital-truncation-rods provides quantitative
information on the interfacial roughness (interfacial
width) and the in-plane correlations at the orbital surface.

\begin{figure}
\centering
\includegraphics[width=0.9\columnwidth]{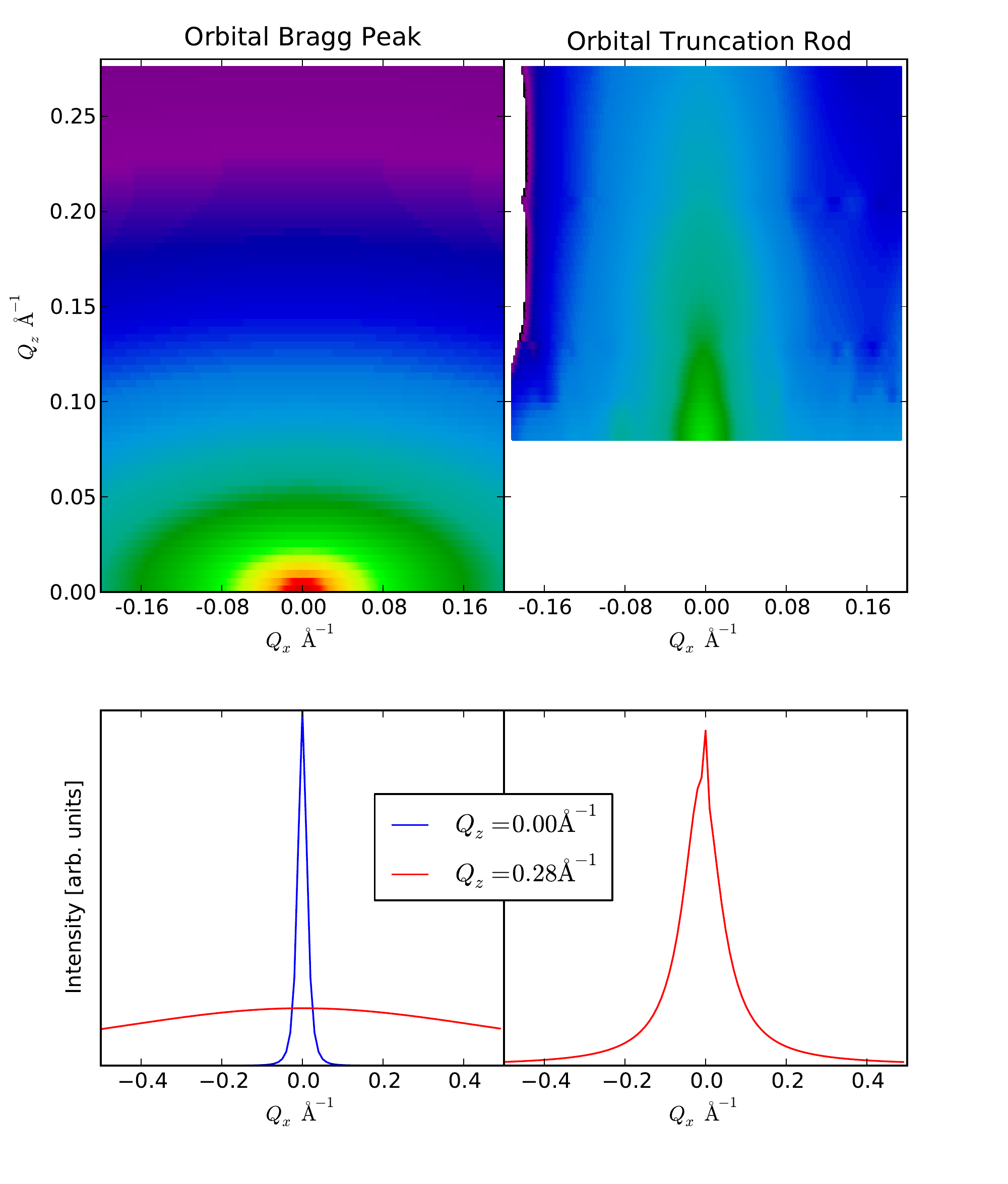}
\caption{(color online) (Top left) Simulated x-ray scattering in the (Q$_{x}$,Q$_{z}$)
plane for a 2D Lorentzian-squared function, with
widths equal to those of the bulk orbital order reflection.(Bottom
left) Line cuts at Q$_{z}$ = 0 {\AA}$^{-1}$ (blue) and Q$_{z}$ = 0.28
{\AA}$^{-1}$ (red) through the simulated x-ray scattering intensity.  (Top
right) 2D Intensity plot of the measured intensity for the orbital
truncation rod in La$_{0.5}$Sr$_{1.5}$MnO$_{4}$ at 170 K. 
The streak of scattering, elongated along $L$ due to the truncation of
the orbital order by its surface can clearly be seen.  (Bottom
right) Line cut at Q$_{z}$ = 0.28 {\AA}$^{-1}$ through the orbital
truncation rod. }
\label{fig:fig1}
\end{figure}

The experiments were carried out on beamline 6ID-B at the Advanced
Photon Source. The incident photon energy was 12~keV. A
single crystal of La$_{0.5}$Sr$_{1.5}$MnO$_{4}$, grown in a floating zone
furnace, was post-cleaved in air to reveal a [001] surface
and immediately placed into the vacuum of a closed-cycle
refrigerator. Q$_{z}$ is
along the surface normal. Scans were performed using a 4+2
diffractometer\cite{You:1999p11}, accessing to the (2.25, -0.25, L) orbital rod and
the (2.25, -0.25, 2) orbital Bragg peak. An incidence angle of
0.6~$^\circ$ was used for all measurements. The sample was
indexed with $a = b = 3.78$~{\AA} and
$c = 12.4$~{\AA}. Momentum transfer is either denoted as Q$_{x}$,
Q$_{z}$ in units of \AA$^{-1}$, or L, in units of r.l.u., where 1 r.l.u.
= 0.507 \AA$^{-1}$. Measurements of the (2.25, -0.25, L) orbital-truncation rod were
made by scanning along Q$_{x}$, perpendicular to the Q$_{z}$ direction.
A 2-D mesh of the intensity obtained from a series of such scans,
collected at 170 K, is shown in the top right panel of Fig.~\ref{fig:fig1}.
The top left hand side of Fig.~\ref{fig:fig1} shows a simulation of the expected
intensity of the superlattice reflection, in the absence of
the orbital truncation rod, constructed from a 2D Lorentzian-squared 
function with Q$_{z}$ and Q$_{x}$ widths equal to the values measured
at the (2.25, -0.25, 2) bulk superlattice reflection.
The bottom panels show line cuts through the respective meshes.
It is immediately apparent that the two behaviors are quite different;
the observed scattering does not originate from tails of the orbital
Bragg peak, but rather is the orbital truncation rod\cite{Wakabayashi:2007p679}.

\begin{figure}
\centering
\includegraphics[width=0.9\columnwidth]{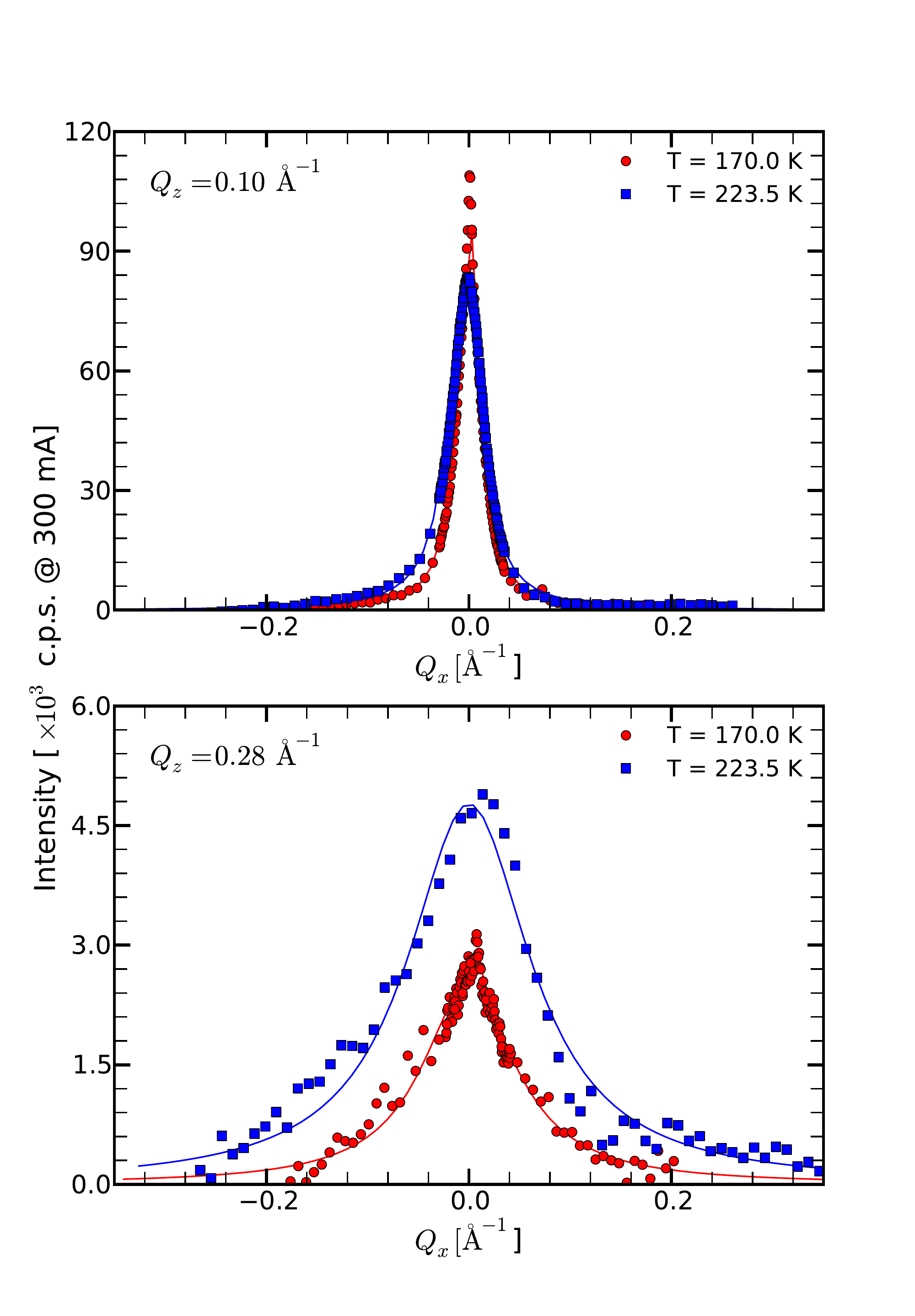}
\caption{(color online) (Top panel) Q$_{x}$ scans
through the orbital truncation rod at 170 K and 223.5 K (red
circles and blue squares respectively). Data taken at Q$_{z}$=
0.1 {\AA}$^{-1}$ (Top panel) and Q$_{z}$ =
0.28 {\AA}$^{-1}$ (Bottom panel). The results of the fits to Eq. 1 are shown as
solid lines.}
\label{fig:fig2}
\end{figure}

Figure~\ref{fig:fig2} shows two Q$_{x}$ scans through this orbital truncation
rod, taken at Q$_{z }$= 0.1~\AA$^{-1}$ and 0.28~\AA$^{-1}$, for two temperatures,
T=170.0 K and 223.5 K, i.e. well below, and just
below, the bulk ordering temperature, T$_{OO}$=226 K. At low temperatures,
a two-component lineshape is observed in the truncation rod scattering.
This immediately implies that the orbital surface has a well-defined
average height but finite in-plane correlations at the surface
\cite{Wakabayashi:2007p679, 0022-3719-18-35-008}. We refer to these two components as the sharp and broad
components. In the
theory of a nearly-smooth surface\cite{0022-3719-18-35-008}, the sharp component results
from the average surface, and the intensity variation of this
component as a function of Q$_{z}$ is a measure of the roughness,
or interfacial width, of that surface. A rough surface results
in a faster decay of the sharp component intensity with Q$_{z}$,
compared to that of a perfectly smooth surface. The broad component
on the other hand, results from the finite in-plane correlation
length of the surface, with a width inversely proportional
to that correlation length. The rate of increase of the integrated
intensity of the broad component as a function of Q$_{z}$ increases
with increasing interfacial roughness. We find that the two length-scale
scattering is best fit by a Lorentzian-squared function for the
sharp component and a Lorentizan function for the broad component:
\begin{equation}
I_T = I_S\left[ \frac{\Gamma^2_S}{Q^2_x + \Gamma^2_S}\right ]^2 + I_B\left[ \frac{\Gamma^2_B}{Q^2_x + \Gamma^2_B}\right ]
\end{equation}
where I$_{S}$ and I$_{B}$ are the peak intensities of the sharp and
broad components respectively, and $\Gamma_{S}$ and $\Gamma_{B}$ are
the respective
widths. Correlation lengths $\xi$, are defined as $\xi = 1 / \Gamma$. For a perfect crystal, the presence of an average surface
would give rise to a delta-function sharp component\cite{0022-3719-18-35-008}. For
the present case, with a finite bulk correlation length, we model
the sharp component with a Lorentzian-squared function\cite{Wakabayashi:2007p679} and
constrain the width of this component to be equal to the width
of the bulk orbital order superlattice reflection, $\Gamma_{S }$= 0.0055
\AA$^{-1}$.

Taking a close look at the T=170 K data, we see that at large
Q$_{z}$, the broad component has a much greater integrated intensity
than the sharp component, while at small Q$_{z}$, the sharp component
dominates. This is the expected behavior for a nearly smooth
surface. Note that while the broad component is much broader
than the sharp component, it is not as broad as would be expected
if this was a line cut through the tails of the orbital Bragg
peak (bottom panel, Fig.~\ref{fig:fig1}), that is both the sharp and broad
components are associated with the orbital truncation rod, and
hence reflect the properties of the orbital surface. At low
temperatures, we find that the in-plane correlation length of
the orbital surface is $1/\Gamma_{B}$=13~\AA, significantly shorter
than the bulk correlation length ($1/\Gamma_{s}$=180~\AA), and in agreement
with earlier work\cite{Wakabayashi:2007p679}.

\begin{figure}
\centering
\includegraphics[width=0.5\columnwidth]{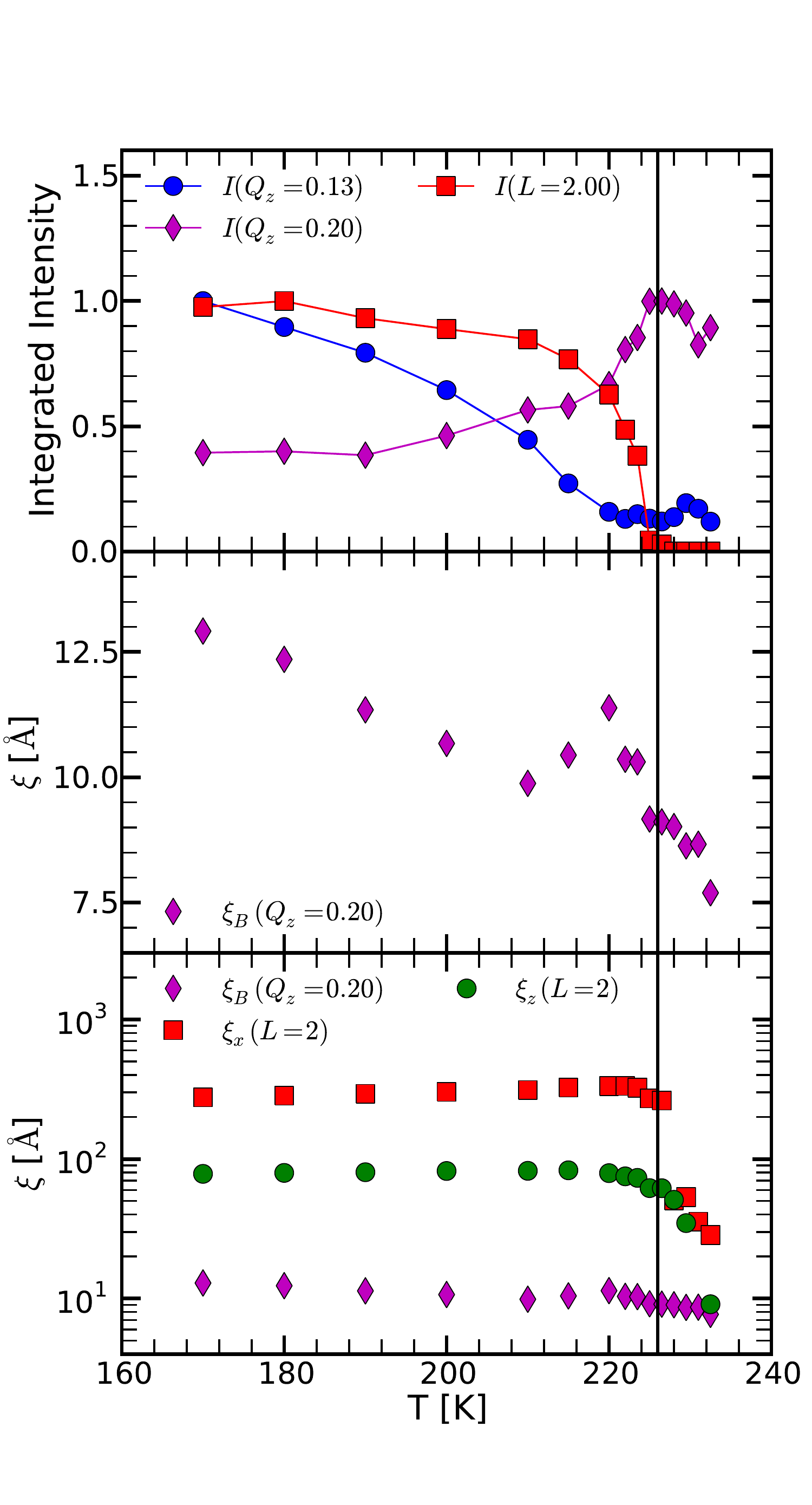}
\caption{(color online) (Top panel) Temperature dependence of the integrated
intensity of the bulk orbital order superlattice reflection (red
squares) measured at $L = 2$ , the integrated intensity
of the sharp component of the orbital truncation rod measured
at Q$_{z}$ = 0.13{\AA}$^{-1}$ (blue circles) and the broad component measured
at Q$_{z}$ = 0.2 {\AA}$^{-1}$ (magenta diamonds). (Middle panel) Temperature
dependence of the correlation length as measued on the the broad component of the
orbital truncation rod at Q$_{z}$ = 0.2 {\AA}$^{-1}$. (Bottom
panel) Temperature dependence of the correlation lengths of the bulk orbital
order superlattice reflection measured along the Q$_{z}$ and Q$_{x}$ directions
(green circles and red squares respectively) and the correlation
length as measured at the on the broad component of the orbital truncation 
rod at Q$_{z}$ = 0.2 {\AA}$^{-1}$.}
\label{fig:fig3}
\end{figure}

Having established that we are probing the orbital surface,
we can now study the temperature dependence of that surface
upon warming through the bulk orbital order transition. The T
= 223.5 K data of Fig.~\ref{fig:fig2} already reveal dramatic changes from
the low temperature lineshape. Specifically, there is no longer any evidence
of a sharp component at any Q$_{z}$, and the broad component is
both broader and more intense than it was at low temperatures.

Figure~\ref{fig:fig3} shows the evolution of this behavior in detail. In the
top panel, the temperature dependence of the integrated intensity
of the bulk superlattice Bragg peak (L = 2), the sharp component
of the surface scattering, measured at Q$_{z}$ = 0.13 {\AA}$^{-1}$, and
the broad component as measured at Q$_{z}$ = 0.2 {\AA}$^{-1}$, are shown.
As the bulk orbital ordering transition is approached, the amplitude
of the bulk superlattice peak decreases, disappearing at T$_{OO}$
= 226 K. The integrated intensity of the sharp component falls
faster than the bulk orbital order Bragg peak and becomes indistinguishable
from zero at T=222~K, i.e below T$_{OO}$. In contrast, the integrated
intensity of the broad component smoothly \emph{increases}. The
correlation length of the bulk orbital order is also shown, measured
along the Q$_{z}$ and Q$_{x}$ directions, in the bottom panel of Fig.~\ref{fig:fig3},
along with the correlation length as measured on the the broad component, along the
Q$_{x}$ direction. The latter is also plotted on a linear scale
in the middle panel of Fig.~\ref{fig:fig3}. These data show that at low temperatures,
the bulk orbital order is anisotropic, with the
larger correlation length in the Q$_{x}$ direction. This correlation
length is larger than that of the broad component, indicative
of a larger in-plane correlation length in the bulk compared
to the orbital surface. As the temperature is raised, the correlation length of
the bulk orbital reflection remains constant until T$_{OO}$ at which
point it rapidly decreases, becoming isotropic above the transition.
In stark contrast, the corrleation length as measured on the broad
component decreases
steadily as the temperature is raised (middle panel, Fig.~\ref{fig:fig3}),
indicative of a slow reduction of the in-plane interfacial correlation
length. Above T$_{OO}$, the correlation lengths of the bulk orbital order Bragg
reflection and the broad component become almost equal.

Before discussing the significance of these experimental results,
it is useful to first discuss what the possible behavior for
the orbital surface might be, on approaching the bulk orbital
order phase transition. One possibility is that of \emph{surface
freezing}, in which an ordered region exists at the surface of
a disordered bulk. Soft matter systems such as liquid crystals\cite{Swanson:1989p2115}
and alkanes\cite{Wu:1993p2062,Ocko:1997p2104} exhibit such behavior. A second, more common, scenario is that of surface melting (or pre-melting).
Surface melting manifests itself as a disordered, liquid-like,
region on the surface of a material, which grows in thickness
upon approaching the bulk transition temperature from below\cite{Frenken:1985p2045}.
In principle, electronic surfaces
could exhibit either of these behaviors. In the study of magnetic surfaces, for example,
there have been various experimental observations of differing
surface and bulk magnetic phase transitions \cite{Binder:1974p2065,Mills:1971p2066}. On the theoretical
side, it was found that for the ferromagnetic Ising model, -
a model that has been used for orbital order in the manganites\cite{Moreo:2000p2053} - it is possible to obtain either surface melting \emph{or}
surface freezing behavior, depending on the ratio of the exchange
interactions for the surface and the bulk\cite{TSALLIS:1988p2067}. The question
is then, which of these scenarios is the relevant one for
the orbital surface discussed here?

\begin{figure}
\centering
\includegraphics[width=0.9\columnwidth]{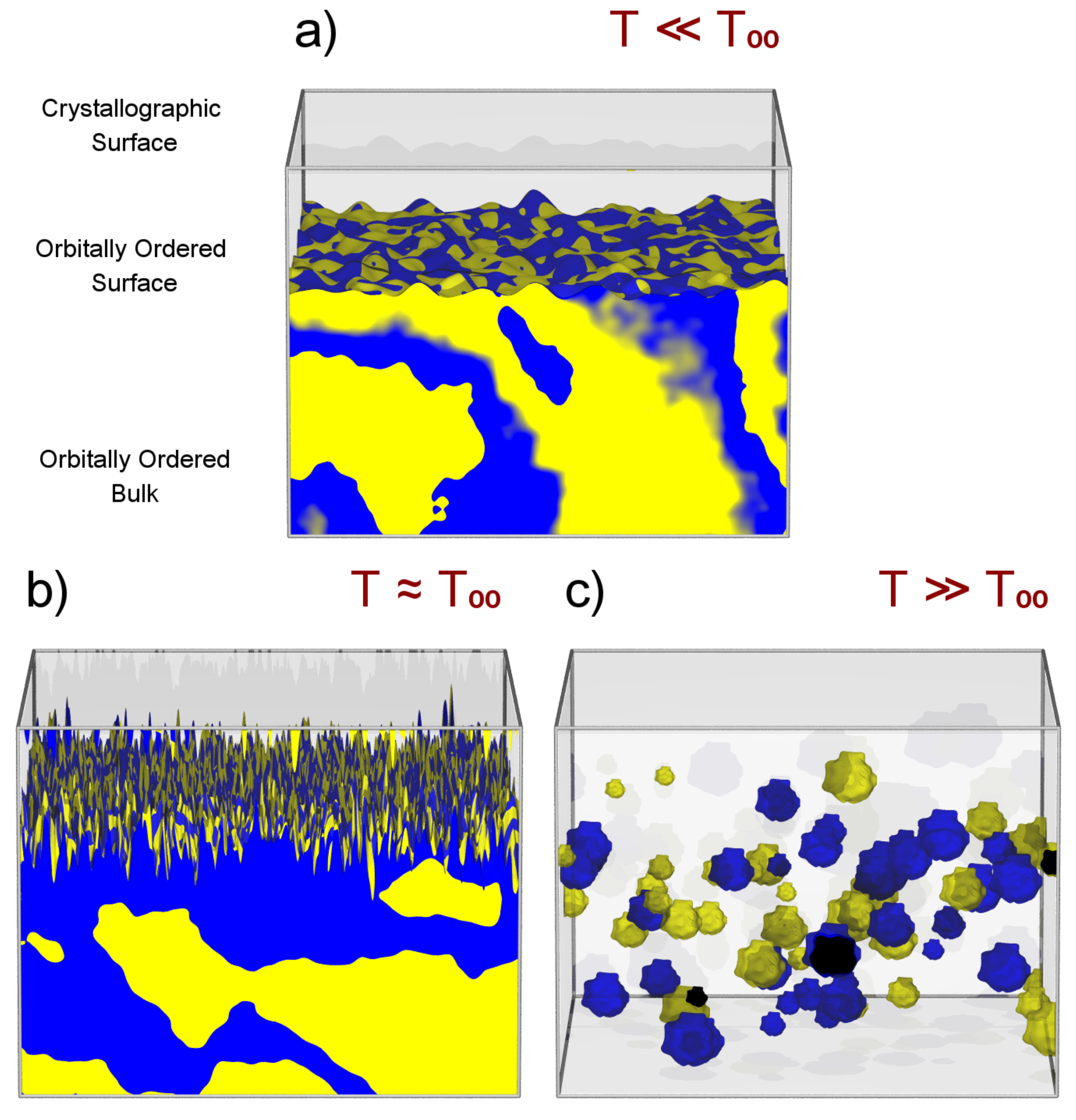}
\caption{(color online) Schematic representation of the
  crystalographic surface, the orbital surface and the orbitally
  ordered bulk for 
(a) $ T \ll T_{oo}$, (b) $T \approx T_{oo}$ and (c) $T
\gg T_{oo}$. The size of the blue and yellow regions in the orbitally
orderd bulk represent schematically the
correlation length of the orbital order in the bulk. The roughness of the orbital surface is shown as the
height variations of the shaded (blue and yellow) surface and the in-place correlation
length by the size of the shaded surface regions.  The size of the all
the colored regions are physically correct, 
both relative to each temperature snapshot and are consistent with the x-ray scattering data analysis.}
\label{fig:fig4}
\end{figure}

The data presented above reveal very different behavior for the
electronic surface and the bulk orbital order on warming through
the bulk orbital ordering transition: At T = 170 K, well below
the bulk transition temperature, the orbital surface is well
defined but rougher than the crystal chemical surface, and has
a shorter correlation length than the bulk orbital order. This
situation is shown pictorially in Fig.~\ref{fig:fig4}(a) where we
reprosent the different correlation lengths by regions of blue and
yellow shading. 

As the temperature is increased, the integrated intensity of
the sharp component falls and the integrated intensity of the
broad component increases, indicating that the interfacial width,
or roughness, is increasing. Further,
the correlation length measured on the broad component is found to steadily decrease,
in contrast to the correlation length of the bulk orbital order, which remains
constant throughout this temperature regime. This implies that
the in-plane correlation length at the orbital surface is decreasing,
while the bulk correlation length is unchanged.

At a temperature just below the bulk orbital order phase transition,
the integrated intensity of the sharp component has dropped below
detectable levels, and the broad component completely dominates,
indicating that there is no average orbital surface at this temperature,
and that the orbital surface has melted. This is shown pictorially
in Fig.~\ref{fig:fig4}(b). It is important to note that at this temperature,
the bulk is still
as well correlated as it was at low temperatures: Thus, this
is a case of electronic surface melting.

Increasing the temperature further, one crosses the bulk phase
transition and the bulk correlation lengths decrease to have
approximately the same magnitude as the interfacial correlation
length. In the absence of an surface, this implies that the
orbital order exists as fluctuations in a disordered background.
Further increasing the temperature causes the size of these fluctuations
to decrease along with the integrated intensity, until they are
no longer observable. This case is shown in Fig.~\ref{fig:fig4}(c).

We conclude by noting that this observation of electronic surface melting has potentially
profound implications on the physics of strongly correlated materials
and their applications in any device. Our results show that,
at least for this system, the electronic surface becomes increasingly
rough and less well defined on warming, and vanishes below the
bulk ordering temperature. Materials with electronic surfaces
that behave in this way could be problematic in devices
operating at finite temperature. For example, colossal magnetoresistance\cite{Jin:1994p2056}
is a potentially very useful property, offering a dramatic change
in resistance in applied magnetic fields. However, in manganites,
this effect requires the system to operate at a temperature close
to the bulk transition temperature. The present work shows that
this temperature corresponds to the roughest (thickest) orbital surface,
which could be a problem for electron transport across that surface.
Further, the properties of any nano-scaled device are likely
to be dominated by the surface / interface behavior due to the
large surface-to-volume ratio. Continuing with the CMR example,
the stark difference between the surface and bulk at this temperature
means that at the nanoscale, CMR may
manifest in a very different form - if at all. Finally, we comment
that while the data presented here applies only to the orbital
surface of LSMO, the implications of this work are likely to
extend much further. For example, we note that much electronic information
about strongly correlated systems is obtained from inherently
surface sensitive techniques such as STM and ARPES. For systems
that exhibit behavior analogous to the present case of electronic
surface melting, such techniques would measure the melted, rough
layer, and not reveal the true bulk temperature evolution. Our
results suggest that future experimental and theoretical investigations
of the temperature dependence of electronic surfaces are to
be encouraged.

The work at Brookhaven was supported by the U.S. Department of Energy,
Division of Materials Science, under Contract No. DE-AC02-98CH10886.
Use of the Advanced Photon Source was supported by the U.S. Department
of Energy, Basic Energy Sciences, Office of Science, under Contract
No. W-31-109-Eng-38. Work in Japan was supported by the Ministry of
Education, Culture, Sports, Science and Technology of Japan (KAKENHI 21740274,19052002)

\bibliography{papers}
\end{document}